\documentclass[12pt,eqsecnum,amsfonts,amssymb]{revtex4}

\usepackage[usenames]{color}
\usepackage{graphicx}
\usepackage{bm}
\usepackage{rotating}

\newcommand{\beq}{\begin{equation}}
\newcommand{\eeq}{\end{equation}}
\newcommand{\beqs}{\begin{eqnarray}}
\newcommand{\eeqs}{\end{eqnarray}}

\begin{document}

\baselineskip 6.0mm

\title{Jones Polynomials and their Zeros for a Family of Knots and Links}

\author{Yue Chen and Robert Shrock}

\affiliation{C. N. Yang Institute for Theoretical Physics and
Department of Physics and Astronomy, \\
Stony Brook University, Stony Brook, New York 11794, USA }

\begin{abstract}

We calculate Jones polynomials $V(H_r,t)$ for a family of alternating
knots and links $H_r$ with arbitrarily many crossings $r$, by
computing the Tutte polynomials $T(G_+(H_r),x,y)$ for the associated
graphs $G_+(H_r)$ and evaluating these with $x=-t$ and $y=-1/t$.  Our
method enables us to circumvent the generic feature that the
computational complexity of $V(L_r,t)$ for a knot or link $L_r$ for
generic $t$ grows exponentially rapidly with $r$. We also study the
accumulation set of the zeros of these polynomials in the limit of
infinitely many crossings, $r \to \infty$.

\end{abstract}

\maketitle

\newpage

\pagestyle{plain}
\pagenumbering{arabic}
\renewcommand{\thefootnote}{\arabic{footnote}}
\setcounter{footnote}{0}


\section{Introduction}
\label{intro_section}

The properties of knots and links are of fundamental interest in
mathematics and mathematical physics \cite{rolfsen}-\cite{ww}.  Here a
link is defined as an embedding of $n_c$ closed curves, each
homeomorphic to $S^1$, in ${\mathbb R}^3$, and a knot is the special
case $n_c=1$. We shall generally use the notation $L$ and $K$ for
links and knots and $L$ when referring to both.  A longstanding goal
has been to achieve a classification of links, including a method for
distinguishing different links.  Progress toward this goal has
involved the construction of various different algebraic formulations
of knots and links, specifically certain polynomials.  Here we shall
focus on the Jones polynomial, $V(L,t)$ \cite{jones85,jones87}.
Technically, this is not, in general, a polynomial, since it can
involve non-integral powers of its variable $t$ and, even in cases
where it only involves integral powers, these may be negative, so that
it is a Laurent polynomial.  However, $V(L,t)$ can always be written
as a polynomial in $t$ multiplied by a monomial prefactor.  With this
qualification, we shall follow common usage in referring to $V(L,t)$
as a (Laurent) polynomial. The usual pictorial representation of knots
and links involves a projection from the embedding space ${\mathbb
  R}^3$ to ${\mathbb R}^2$ with a number $r$ of resultant apparent
crossings of the curves on one or more entangled $S^1$ circuits.  A
standard list of knots and links was given by Rolfsen \cite{rolfsen},
and we follow the labelling conventions in \cite{rolfsen}.  According
to these labelling conventions, the symbol $r^{n_c}_j$ denotes the
$j$'th link with $n_c$ different $S^1$ components and $r$ crossings.
Thus, for example, $6^2_3$ refers to the link with $n_c=2$ different
$S^1$ circuits entangled with $r=6$ crossings, ordered as the third
member in the list of such links, and $6_2 \equiv 6^1_2$ refers to the
knot with $r=6$ crossings, ordered as the second member of the list of
such knots, etc. (Note that, as indicated, if $n_c=1$, then the
superscript is omitted in this notation.)  We recall that an
alternating link is defined as a link in which, as one travels along
each of the $n_c$ entangled $S^1$ circuits, one traverses a crossing
in an overlying manner, then underlying, then overlying, and so
forth. To each such alternating link $L$, one can assign an associated
graph, $G_+(L)$, as will be discussed below.

Given the usefulness of the Jones polynomial in distinguishing
different knots and links, an early question that arose concerned the
question of the calculational complexity of this polynomial. It was
shown by Jaeger, Vertigan, and Welsh \cite{jvw} that the computational
complexity of the Jones polynomial $V(L,t)$, for general $t$, is
$\sharp{\rm P}$ hard. Roughly speaking, this means that the difficulty
of this calculation increases exponentially rapidly as a function of
the number of crossings, $r$. (Exceptions occur at the special points
$t=\pm 1$, $t=\pm i$, $t=e^{\pm 2 i \pi/3}$, and $t=e^{\pm 4i\pi/3}$
\cite{jvw}). For a general knot or link and a general value of $t$,
this renders the calculation of the Jones polynomial intractable as
the number of crossings becomes sufficiently large. One is then
naturally led to inquire whether there is a way to circumvent this
intractability, at least for some sets of knots and links.  In
\cite{jz}, Chang and Shrock presented a method that achieves this goal
and applied it to several infinite families of alternating knots and
links.  Detailed definitions will be given in the next section, but
here we briefly describe this method. One determines the associated
graphs for two or more specific knots or links, and generalizes these
to an infinite set of graphs $G_m$ with a recursive structure, which
can thus be labelled with an index $m$, where $m$ is a function of
$r$.  The number of crossings on the knot or link is equal to the
number of edges of the associated graph.  One then calculates a
general formula for the Tutte polynomial $T(G_m,x,y)$ and takes the
special case $x=-t$ and $y=-1/t$ to obtain the corresponding Jones
polynomial for a knot or link in this class with arbitrarily many
crossings.  For this infinite class, the method presented in \cite{jz}
thus circumvents the generic exponential increase in the difficulty of
calculation of Jones polynomials of knots and links with the number of
crossings. Ref. \cite{jz} also studied the zeros of several infinite
families of knots and links, including their respective accumulation
sets in the limit of infinitely many crossings, $r \to \infty$. In
addition to \cite{ww} and \cite{jz}, some later studies of zeros of
Jones polynomials were reported in
\cite{jin_zhang2003}-\cite{dong_jin2015}. In passing, we mention that,
in addition to Jones polynomials, there are a number of others that
have been studied to gain information about knots and links, including
Alexander, Conway, HOMFLY, Kauffman, and Akutsu-Wadati polynomials, as
reviewed, e.g., in \cite{kauffman}-\cite{bollobas}.  However, we
restrict our consideration here to Jones polynomials.

In the present paper we will apply the method of \cite{jz} to
calculate Jones polynomials for a family of alternating knots and
links for which this calculation is somewhat more complicated, in the
sense that the Jones polynomial of the $m$'th member of the family
involves powers of a larger number of terms than those considered in
\cite{jz} (see Eqs. (\ref{vlsum}), (\ref{Nlambda_jz}), and
(\ref{Nlambda_hm}) below).  We will also discuss some properties of
the zeros of these Jones polynomials, including the locus of zeros in
the limit $r \to \infty$.  In order to describe our methods, it is
first necessary to review some relevant background concerning knots
and links, graph theory, and Jones and Tutte polynomials. We turn to
this next.


\section{Background} 
\label{background_section}

\subsection{Connection between Jones and Tutte Polynomials} 

Here we review some background on properties of knots and links and
the calculation of Jones polynomials, including the connection with
Tutte polynomials.  Further details may be found in
\cite{rolfsen}-\cite{bollobas}.  For a given knot or link $L$, we
first define what are known as the corresponding ``shaded-region''
diagrams.  Let us denote a crossing on the knot as $- {\Big |} -$,
where, by convention, we orient the over-going strand as vertical (and
the apparent break in the lower strand indicates that it is occluded
by the upper strand). In the shaded-region diagram $D_+(L)$, for each
such crossing, one shades the upper left and lower right regions
adjacent to this crossing so that they are dark, while the upper right
and lower left regions are unshaded (i.e, light). Similarly, in the
shaded-region diagram $D_-(L)$, for this crossing $- {\Big | } -$, one
shades the upper right and lower left regions adjacent to the
crossing, and leaves the upper left and lower right regions
unshaded. This assignment is independent of the direction of motion
along the given $S^1$ curve on $L$. We denote the numbers of dark
($d$) and light ($\ell$) regions in the shaded region diagram as
$n_d(D_\pm(L))$ and $n_\ell(D_\pm(L))$.  It is easily seen that
$n_d(D_+(L))=n_\ell(D_-(L))$ and $n_\ell(D_+(L))=n_d(D_-(L))$.  One
then constructs the associated graphs $G_+(L)$ and $G_-(L)$ by
assigning vertices to the shaded regions and edges (bonds) connecting
these vertices by traversing the adjacent crossings.  Without loss of
information, one may choose to deal only with $D_+(L)$ and $G_+(L)$ or
$D_-(L)$ and $G_-(L)$. Here we choose $D_+(L)$ and $G_+(L)$.

A graph $G=(V,E)$ is defined by its vertex and edge sets $V$
and $E$.  Two vertices connected by an edge are said to be adjacent.
A general graph may contain multiple edges joining two (adjacent)
vertices and may also contain one or more edges that connect a vertex
back to itself. Let the number of vertices and edges of the graph $G$
be denoted $n(G)=|V|$ and $e(G)=|E|$. Clearly, the total number of
crossings is equal to the number of edges of the associated graph:
\beq
r(L)=e(G_+(L))=e(G_-(L)) \ ,
\label{crossing_eq_edges}
\eeq
and the respective numbers of dark (light) regions in the $D_+$ ($D_-$) 
region diagrams is equal to the number of vertices in the associated graphs:
\beq
n_d(D_+)=n(G_+(L)), \quad n_d(D_-)=n(G_-(L)) \ .
\label{dark_regions_eq_vertices}
\eeq
Furthermore, $G_-(L)$ is the (planar) dual of $G_+(L)$, i.e., 
\beq
G_+(L)=[G_-(L)]^* \ .
\label{gdual}
\eeq
Here, we recall a definition: given a planar graph $G$, the (planar)
dual graph $G^*$ is constructed by assigning to each face of $G$ a
vertex in $G^*$ and connecting these vertices by edges
traversing the edges of $G$.

Another property of a knot or link is its ``writhe'', $w(L)$. To
define this, we start by associating a $\pm$ sign to each crossing, as
follows.  let us choose a direction of motion along each $S^1$ circuit
in a knot $L$, and, by convention, take the direction of motion on the
overlying strip at a crossing as being from bottom to top. Then if the
motion on the underlying strand is from left to right, assign a minus sign
$s_-$ to this crossing, and if it is from right to left, assign a plus sign
$s_+$ sign to the crossing. Pictorially,
$\leftarrow {\Big \uparrow} \leftarrow$ is assigned a $s_+$ sign, 
while $\rightarrow {\Big  \uparrow}\rightarrow$ is assigned a $s_-$ sign. 
For a given knot or link $L$, the numbers of the $\pm$ crossings are
denoted $N_{s_+}(L)$ and $N_{s_-}(L)$, satisfying
$N_{s_+}(L)+N_{s_-}(L)=r(L)$.  Then the writhe of $L$ is given by
\beq
w(L) = N_{s_+}(L)-N_{s_-}(L) \ . 
\label{writhe}
\eeq

To each knot or link $L$, there corresponds another, which we
label as $R(L)$, obtained by replacing each 
over-under crossing by an under-over crossing and vice versa.  A basic
property of the Jones polynomial under this reversal procedure is that
\beq
V(L,t)=V(R(L),t^{-1}) \ .
\label{vlvrl}
\eeq

We next proceed to discuss the relation between the Jones polynomial
$V(L,t)$ and the Tutte polynomial of the associated graph $G_+(L)$
which applies for the case in which $L$ is an alternating link or
knot.  The Tutte polynomial $T(G,x,y)$ of a graph $G$
is \cite{tutte1}-\cite{tutte2}, \cite{welsh,bollobas}
\beq
T(G,x,y)=\sum_{G^\prime \subseteq G} (x-1)^{k(G^\prime)-k(G)}
(y-1)^{c(G^\prime)} \ ,
\label{tuttepol}
\eeq
where $G^\prime$ is a spanning subgraph of $G$, i.e., $G^\prime$ has
the same vertex set of $G$ and a subset of the edge set of $G$:
$G^\prime = (V,E^\prime)$ with $E^\prime \subseteq E$.  In
eq. (\ref{tuttepol}), $k(G^\prime)$, $e(G^\prime)=|E^\prime|$, and
$c(G^\prime)$ denote the number of components, edges, and (linearly
independent) circuits of $G^\prime$, with the usual relation
\beq
c(G^\prime) = e(G^\prime)+k(G^\prime)-n(G^\prime) \ , 
\label{ceq}
\eeq
where, as above, $n(G^\prime)$ is the number of vertices of $G^\prime$. 
The Tutte polynomials of a planar graph $G$ and its (planar) dual $G^*$ are
related according to
\beq
T(G,x,y)=T(G^*,y,x) \ .
\label{ttdual}
\eeq

The relation between the Jones polynomial $V(L,t)$ and the Tutte polynomial 
of the associated graph $G_+(L)$ for an alternating link or knot is then 
\beq
V(L,t) = (-1)^{w(L)} \, t^{p_t(L)} \, T(G_+(L),-t,-1/t) \ , 
\label{vtrel}
\eeq
where the power $p_t(L)$ is 
\beq
p_t(L) = \frac{1}{4}\bigg [ n_\ell(G_+(L))-n_d(G_+(L))+3w(L) \bigg ] \ .
\label{pt}
\eeq
For a link $L$ involving an odd number $n_c$ of $S^1$ components and
hence, in particular, for a knot, $p_t(L)$ is an integer, so $V(L,t)$ is
a Laurent polynomial in $t$. For a link consisting of an even number
$n_c \ge 2$ of components $n_\ell(G_+(L))-n_d(G_+(L))+3w(L) = 2$ mod 4, so
$p_t(L)$ is a half-odd-integer.

We remark on two basic properties of the zeros of Jones polynomials.
First, since the coefficients of the terms in the Jones polynomial are
real, it follows that these zeros are either real or are comprised of
complex-conjugate pairs, so that the set of zeros is invariant under
complex conjugation.  Second, since terms in $V(L,t)$ with successive
powers of $t$ have alternating signs, it follows that $V(L,t)$
has no zeros on the negative real axis in the $t$ plane (e.g.,
\cite{dong_jin2015}).

A brief review of a general method for calculating a Tutte polynomial
is in order here.  Consider a graph $G=(V,E)$ with vertex and edge
sets $V$ and $E$, and let $e$ be an edge in the set $E$. Let $G-e$
denote the graph obtained from $G$ by deleting the edge $e$ and let
$G/e$ denote the graph obtained from $G$ by deleting the edge $e$ and
identifying the vertices that were connected by $e$.  An edge $e=e_b$
is a bridge if its deletion increases the number of components in $G$.
An edge $e=e_\ell$ is a loop if it connects a vertex back to itself.
If a graph has no edges, then the Tutte polynomial of this graph is
unity.  In general, the Tutte polynomial $T(G,x,y)$ satisfies the
following properties:

\begin{enumerate}

\item
  If an edge $e_b \in E$ is a bridge, then
\beq
T(G,x,y)=x T(G-e_b,x,y) \ . 
\label{tbrel}
\eeq

\item
  If an edge $e_\ell \in E$ is a loop, then
\beq
T(G,x,y) = yT(G/e_\ell,x,y) \ . 
\label{tellrel}
\eeq

\item
  If an edge $e \in E$ is neither a bridge nor a loop, then $T(G,x,y)$
satisfies the deletion-contraction relation 
\beq  
T(G,x,y)=T(G-e,x,y)+T(G/e,x,y) \ .
\label{dcr}
\eeq

\end{enumerate}
%


\subsection{Some General Properties of Jones Polynomials}

Two basic properties of Jones polynomials concern the highest and
lowest powers of $t$ in $V(L,t)$ for an alternating knot or link.  We
first note the values of the respective maximal degrees of the Tutte
polynomial in its variables $x$ and $y$:
\beq
{\rm max}({\rm deg}_x[T(G,x,y)]) = n(G)-1
\label{maxdegx}
\eeq
and
\beq
{\rm max}({\rm deg}_y[T(G,x,y)]) = c(G) \ .
\label{maxdegy}
\eeq
Then, combining (\ref{maxdegx}) and (\ref{maxdegy}) with
(\ref{vtrel}), we have, for an alternating link
\beq
{\rm max}({\rm deg}_t(V(L,t)) =  n(G_+(L))-1 + p_t(L)
\label{hipowvl}
\eeq
\beq
{\rm min}({\rm deg}_t(V(L,t)) = -c(G_+(L))+ p_t(L) \ . 
\label{lopowvl}
\eeq
 From these results, together with the
relation (\ref{ceq}), it follows that 
\beq
    {\rm max}({\rm deg}_t(V(L,t)) -
    {\rm min}({\rm deg}_t(V(L,t)) = e(G_+(L)) = r(L) \ .
\label{degdif}
\eeq
%


\section{Calculational Method}
\label{method_section}

In this section we discuss in detail the calculational method presented in
\cite{jz}, which we use here. This method consists of the following
steps: 

\begin{enumerate}

\item

  First, one determines the associated graphs for two or more specific
  knots or links, and establishes that these graphs form members of an
  infinite set of graphs with a recursive structure. Let us denote the
  $m$'th such associated graph as $G_m$, where $m$ is a function of
  $r$, the number of crossings on the given knot or link.  Here and
  elsewhere, we sometimes explicitly indicate this functional
  dependence $m=m(r)$, or equivalently, $r=r(m)$, but often leave it
  implicit in the notation, writing $L_{r(m)} \equiv L_r$ and
  $G_{m(r)} \equiv G_m$. The functions $r(m)$ and equivalently $m(r)$ depend
  on the particular family of knots and links (see Eq. (\ref{rmrel}) for
  the family under consideration here). By the term
  ``recursive'', one means that the graph $G_{m+1}$ can be obtained
  from the graph $G_m$ by a systematic procedure involving addition of
  vertices and edges; an example is a ladder graph of length $m$
  squares.  The number of vertices of $G_m$ is equal to the number of
  dark regions in the shaded diagram of the knot or link, and the
  number of edges of $G_m$ is equal to the number of crossings in each
  knot or link, as given by Eqs. (\ref{dark_regions_eq_vertices}) and
  (\ref{crossing_eq_edges}).

\item
  Next, one uses the deletion-contraction relation iteratively to
  calculate a general formula for the Tutte polynomial $T(G_m,x,y)$
  that is valid for arbitrary $m$. This is a tractable calculation
  because of the recursive nature of the graphs $G_m$. 

  \item
    Using the relation (\ref{vtrel}), one then evaluates $T(G_m,x,y)$ with
    $x=-t$ and $y=-1/t$ to calculate a general expression for a knot
    or link in the given family for arbitrary $m$ and thus arbitrarily
    many crossings, $r$. 

\end{enumerate}

Evidently, this method thus circumvents the problem of exponentially growing
computational complexity of Jones polynomials for each family of knots
studied via this method.  In \cite{jz}, this procedure was applied to the
following knots and links:

\begin{enumerate}

\item
  The family $A_m$ with $G_+(3_1)=D1C_2$, $G_+(4_1)=D1C_3$, and
  $G_+(5_2)=D1C_4$, where $C_m$ denotes the circuit graph with $m$
  vertices and $D1C_m$ is the graph obtained by doubling one of the
  edges of $C_m$. These specific correspondences were generalized to
  a family of knots $A_m$ with $G_+(A_m)=D1C_{m-1}$. 

\item
  The family $B_m$ with $G_+(4_1)=(Wh)_3=D1C_3$, $G_+(6^3_2)=(Wh)_4$, and
  $G_+(8_{18})=(Wh)_5$, where $(Wh)_m$ is the $m$-vertex wheel graph obtained
  from the circuit graph $C_{m-1}$ by connecting each of the $m-1$ vertices
  of $C_{m-1}$ via edges (``spokes'') to another vertex (the ``axle'' 
  of the wheel). These specific correspondences were generalized to a family
  of links such that $G_+(B_m) = (Wh)_m$. 

\item
  The family $E_m$ with $G_-(6^3_1)=DC_3$ and $G_-(8^4_1)=DC_4$, where
  $DC_m$ denotes the graph obtained from the $m$-vertex circuit graph
  $C_m$ by doubling each edge. The general family is
  $G_-(E_m)=DC_m$ for $m \ge 3$. 

\item
  The family $F_m$ with $G_+(6_1)=HW_5$, where $HW_m$ is the graph
  obtained from the wheel graph $(Wh)_{(m+1)/2}$ by inserting a vertex
  on each spoke edge of this precursor wheel graph.

\end{enumerate}


\section{A Structural Theorem for $V(L_{r(m)},t)$} 
\label{Vstructure}

The Tutte polynomial for a recursive graph $G_m$ consisting of $m$
repetitions of some basic subgraph can be written in the form \cite{a}
\beq
T(G_m,x,y) = \sum_{j=1}^{N_{T,G,\lambda}} c_{G,j}(\lambda_{G,j})^m \ , 
\label{tgsum}
\eeq
where the coefficients $c_{G,j}$ and the terms $\lambda_{G,j}$ depend
on the type of graph but not on $m$.  We will use the symbols $L
\equiv \{ L_{r(m)} \}$ and $G \equiv \{ G_m \}$ to refer to the family
of knots or links $L_{r(m)}$ and the corresponding family of
associated graphs $G_m = G_\pm(L_{r(m)})$, respectively. From
Eqs. (\ref{tgsum}) and (\ref{vtrel}) for alternating links, it follows
that the Jones polynomial for links $L_{r(m)}$ whose associated graphs
$G_\pm (L_{r(m)})$ are recursive, has the same type of form, namely
\beq
V(L_{r(m)},t) = \sum_{j=1}^{N_{L,\lambda}} c_{L,j}(t) \,
[\lambda_{L,j}(t)]^m \ , 
\label{vlsum}
\eeq
where 

\beq
N_{L,\lambda} \le N_{T,G_+(L),\lambda} \ .
\label{nlnt}
\eeq
The inequality is used here because when one sets $x=-t$ and $y=-1/t$,
some $\lambda_{G,j}(x,y)$'s may become identically equal or may vanish
\cite{jz}.  It was shown in \cite{jz} that the number of terms
$N_{L,\lambda}$ in the respective realizations of Eq. (\ref{vlsum})
for these families were
\beq
N_{A,\lambda}=2, \quad N_{B,\lambda}=3, \quad N_{E,\lambda}=2,
\quad N_{F,\lambda}=3 \ , 
\label{Nlambda_jz}
\eeq
where, as in Eq. (\ref{vlsum}), the symbol $A$ refers to the family
$\{ A_{m(r)} \}$ in its entirety, and similarly with the other
families of knots and links.  The $H_{r(m)}$ family that we study here
has a more complicated realization of the general structural relation
(\ref{vlsum}), as will be seen in Eq. (\ref{Nlambda_hm}) below.

A term $\lambda_{L,j}(t)$ in (\ref{vlsum}) is dominant at a given point
$t$ in the complex $t$ plane if its magnitude
$|\lambda_{L,j}(t)|$ is greater than the magnitudes of other terms
$\lambda_{L,j^\prime}(t)$ occurring in (\ref{vlsum}).  This definition is
motivated by the fact that in a given region of the complex $t$ plane,
as $m$ gets large, the $[\lambda_{L,j}(t)]^m$ term in (\ref{vlsum}) with
the largest magnitude dominates the sum to a greater and greater
extent. In the limit $r \to \infty$ and thus also $m \to \infty$, the
zeros of $V(L,t)$ in the complex $t$ plane accumulate where two or
more $\lambda$ terms become equal in magnitude, and the limiting
continuous accumulation set of zeros of $V(L,t)$ is determined by the
equality in magnitude of two or more dominant $\lambda$ terms. This is
similar to the case with the continuous accumulation set of zeros of
chromatic polynomials \cite{bkw}.  As in earlier studies of chromatic
and Tutte polynomials \cite{w,a} (equivalent to Potts model partition
functions), we denote this continuous accumulation locus as ${\cal
  B}$.


\section{Family $H_{r(m)}$ with $G_+(H_{r(m)})=S_m$}
\label{hr_section}

As mentioned above, in \cite{jz}, calculations of Jones polynomials
were presented for several families of knots and links such that the
value of $N_{L,\lambda}$ was 2 or 3, as listed in
Eq. (\ref{Nlambda_jz}).  For the two familes where this number was 3, the
respective cubic equation for these roots factorized into a linear and
quadratic factor.  Here we have carried out a study of a family $H =
\{ H_{r(m)} \}$ of knots and links of greater complexity, in the sense
that $N_{H,\lambda}=5$ in Eq. (\ref{vlsum}) and the quintic equation
for the roots factorizes into a linear term times a quartic
polynomial.  This leads to a correspondingly greater complexity in the
pattern of zeros and continuous accumulation locus ${\cal B}$ than was
the case for the families of knots and links studied in \cite{jz}.

From our calculations of the associated graphs of a number of alternating
knots and links, we find that for the knots
$R(6_2)$ and $R(10_{116})$ these graphs are 
\beq
G_+(R(6_2))=S_1
\label{s1}
\eeq
and
\beq
G_+(R(10_{116}))=S_2 \ ,
\label{s2}
\eeq
where $S_m$ is a ladder graph of $m$ squares with free longitudinal
boundary conditions, such that all vertices along one side of the
ladder are connected by edges to a single external vertex. An
illustrative example of an $S_m$ graph with $m=2$ is given by a
$90^\circ$ degree rotation of Fig. 2(c) in \cite{strip}, such that the
transverse direction in that figure becomes the longitudinal
direction. (As was defined above, $R(L)$ is the link obtained from $L$
by reversing the over-under order to under-over order for the strands
at each crossing.)  We generalize this to a family of knots and links
$H_{r(m)}$ with an arbitrarily great number of crossings, satisfying
the property that
\beq
G_+(H_{r(m)}) = S_m \ , 
\label{sm}
\eeq
where the function $r=r(m)$ is  given below in Eq. (\ref{rmrel}). 
The numbers of vertices and edges in the graph $S_m$ are
\beq
n(S_m) = 2m+3
\label{nsm}
\eeq
and
\beq
e(S_m) = 4m+2 \ .
\label{esm}
\eeq
From the general relation ({\ref{crossing_eq_edges}),
the number $r$ of crossings in the $H_r$ knot or link is given by 
\beq
r = r(m) = e(S_m) = 4m+2, \ i.e., \ m = m(r) =  \frac{r-2}{4}  \ . 
\label{rmrel}
\eeq
Thus, as in Eqs. (\ref{s1}) and (\ref{s2}), the realizations of the
lowest-$m$ knots $H_r$ in the Rolfsen list are
\beq
m=1 \ \Leftrightarrow \ r=6: \quad H_6 = 6_2
\label{H6_rolfsen}
\eeq
and
\beq
m=2 \ \Leftrightarrow \ r=10: \quad H_{10} = 10_{116} \ .
\label{H10_rolfsen}
\eeq
Further, for the number of vertices on $S_m$, we have 
\beq
n_d(H_{r(m)}) = n(S_m) = 2m+3 \ .
\label{ndark_eq_nsm}
\eeq
As noted above, we sometimes indicate explicitly the functional dependence
$r=r(m)$ or equivalently $m=m(r)$ but elsewhere leave it implicit in the
notation. 

From calculations of knots and links whose associated graphs are $S_m$
with higher values of $m$ up to $m=10$, we obtain results that are
consistent with the following generalizations:
\beq
n_\ell(H_{r(m)}) = 2m+1 \ ,
\label{nlight}
\eeq
so that
\beq
n_\ell(H_{r(m)}) - n_d(H_{r(m)}) = -2 \ ,  
\label{nlight_minus_ndark}
\eeq
and, for the writhe 
\beq
w(H_{r(m)}) =
\cases{2 & if $m=1, \ 2, \ {\rm or} \ 4$ mod  5 \cr
       6 & if $m=3$ mod 5 \cr
      -2 & if $m=0$ mod 5} 
\label{wm}
\eeq
and hence, combining (\ref{nlight_minus_ndark}) and (\ref{wm}) and
evaluating the power $p_t(L)$ in Eq. (\ref{pt}), 
\beq
p_t(H_{r(m)}) =
\cases{1 & if $m=1, \ 2, \ {\rm or} \ 4$ mod 5 \cr
       4 & if $m=3$ mod 5 \cr
      -2 & if $m=0$ mod 5} 
\label{ptm}
\eeq
With these ingredients, we then can calculate $V(H_{r(m)},t)$ in terms of
$T(S_m,x,y)$ using the evaluation $x=-t$ and $y=-1/t$, via
Eqs. (\ref{vtrel}) and (\ref{rmrel}).

The Tutte polynomials $T(S_m,x,y)$ can be expressed via a generating
function depending on an auxiliary expansion variable, $z$:
\beq
\Gamma(S,x,y,z) = \frac{{\cal N}(S,x,y,z)}{{\cal D}(S,x,y,z)} =
\sum_{m=1}^\infty T(S_m,x,y) z^{m-1} \ .
\label{genfun}
\eeq
By iterative use of the deletion-contraction relation (\ref{dcr}) with
(\ref{tbrel}) and (\ref{tellrel}), we calculate the following
generating function. We find that the degrees of the
numerator and denominator polynomials in $\Gamma(S,x,y,z)$, as functions
of $z$, are
\beq
{\rm deg}_z({\cal N}(S,x,y,z)) = 4
\label{degnum}
\eeq
and
\beq
{\rm deg}_z({\cal D}(S,x,y,z)) = 5 \ . 
\label{degden}
\eeq
Therefore, the numerator and denominator of the generating function
$\Gamma(S,x,y,z)$ have the form 
\beq
{\cal N}(S,x,y,z) = \sum_{j=0}^4 a_j(x,y) z^j
\label{numfun}
\eeq
and
\beq
{\cal D}(S,x,y,z) = 1 + \sum_{j=1}^5 b_j(x,y) z^j \ .
\label{denfun}
\eeq
Here, with no loss of generality, we have normalized the numerator and
denominator polynomials in $\Gamma(S,x,y,z)$ so that the constant term in
${\cal D}(S,x,y,z)$ is unity. 

For the coefficient functions $a_j$ with $0 \le j \le 4$ in
${\cal N}(S,x,y,z)$, we calculate
\beq
a_0 = x^4+2x^3+x^2y+2x^2+2xy+y^2+x+y
\label{a0}
\eeq
\beqs
a_1 &=& -x\bigg [ x^4y+x^3y^2+x^4+4x^3y+2x^2y^2+xy^3+3x^3 \cr\cr
    &+&6x^2y+4xy^2+y^3+3x^2+5xy+3y^2+x+y \bigg ]
\label{a1}
\eeqs
\beqs
a_2 &=& xy\bigg [x^4y^2+x^5+3x^4y+2x^3y^2+4x^4+5x^3y+3x^2y^2 \cr\cr
    &+& 2xy^3+5x^3+5x^2y+3xy^2-y^3+3x^2+2xy-y^2 \bigg ] 
\label{a2}
\eeqs
\beq
a_3 =-x^2y^2\bigg [ x^4y+x^3y^2+x^4+3x^3y+2x^3+2x^2y+2xy^2+x^2-y^2 \bigg ]
\label{a3}
\eeq
\beq
a_4 = x^6y^4 \ .
\label{a4}
\eeq

For the coefficient functions $b_j$ with $j=1,...,5$ in 
${\cal D}(S,x,y,z)$ we calculate
\beq
b_1 = -\bigg [ 3(1+x+y) + xy + x^2+y^2 \bigg ]
\label{b1}
\eeq
\beqs
b_2 &=& 1 + 3(x+y)+3(x^2+y^2)+8xy+x^3+y^3 +5xy(x+y) \cr\cr
    &+&xy(x^2+y^2) +(xy)^2
\label{b2}
\eeqs
\beq
b_3 = -xy\bigg [ 3+5(x+y)+4(x^2+y^2)+6xy+x^3+y^3+3xy(x+y)+(xy)^2 \bigg ]
\label{b3}
\eeq
\beq
b_4 = (xy)^2(1+x)(1+y)(1+x+y)
\label{b4}
\eeq
\beq
b_5 = -(xy)^4 \ . 
\label{b5}
\eeq

Substituting these results for the $a_j$ with $1 \le j \le 4$ and
$b_j$ with $1 \le j \le 5$ in Eq. (\ref{genfun}) with (\ref{numfun})
and (\ref{denfun}) yields the Tutte polynomials $T(S_m,x,y)$.  Then,
setting $x=-t$ and $y=-1/t$, and using Eq. (\ref{vtrel}), we 
obtain the Jones polynomial 
\beq
V(H_{r(m)},t) = (-1)^{w(H_{r(m)})} \, t^{p_t(H_{r(m)})} \, T(S_m,-t,-1/t) \ , 
\label{vhrm}
\eeq
where $w(H_{r(m)})$ and $p_t(H_{r(m)})$ were given in Eqs. (\ref{wm})
and (\ref{ptm}). From Eqs. (\ref{genfun})-(\ref{denfun}) and
(\ref{vhrm}), it follows that $V(H_{r(m)},t)$ has the form of
Eq. (\ref{vlsum}) with
\beq
N_{H,\lambda}=5 \ , 
\label{Nlambda_hm}
\eeq
namely
\beq
V(H_{r(m)},t) = \sum_{j=1}^5 c_{S,j}(t) [\lambda_{S,j}(t)]^m \ . 
\label{vhmsum}
\eeq

From our general formula (\ref{vhrm}) we obtain the known results for the
knots $H_{m=1;r=6} = 6_2$ and $H_{m=2;r=10}=10_{116}$, namely 
\beq
V(R(6_2),t) = t \, T(S_1,-t, -1/t) =
t^5 - 2t^4 + 2t^3 - 2t^2 + 2t - 1 + \frac{1}{t} 
\label{vm1}
\eeq
and
\beqs
V(R(10_{116}),t) &=& t \, T(S_2,-t,-1/t) =
t^7 - 4t^6 + 8t^5 - 12t^4
\cr\cr
&+& 15t^3 - 16t^2
+ 15t - 11 + \frac{8}{t} - \frac{4}{t^2} + \frac{1}{t^3} \ . 
\label{vm2}
\eeqs
The next knot in the series, $H_{r(m)}$ with $m=3$, has $r=r(m)=14$
crossings and has not, to our knowledge, been tabulated.  Denoting it as
above, we calculate
\beqs
V(R(H_{m=3;r=14},t) &=& t^4 \, T(S_3,-t,-1/t) \cr\cr 
&=& t^{12} - 6t^{11} + 18t^{10} - 38t^9 + 64t^8 - 91t^7 + 111t^6 - 118t^5
\cr\cr
&+& 111t^4 - 92t^3 + 66t^2 - 39t + 19 - \frac{6}{t} + \frac{1}{t^2} \ . 
\label{vm3}
\eeqs
Using Eq. (\ref{vhrm}), one can calculate $V(H_{r(m)},t)$ for arbitrarily
large $r$.

It is of interest to remark on a number of properties of these results. 
The Tutte polynomials of a recursive family of graphs satisfy a linear
recursion relation \cite{bds}. In the present case, with the shorthand
notation $T_m \equiv T(S_m,x,y)$ and $b_j \equiv b_j(x,y)$, this is
\beq
T_{m+5}+b_1 T_{m+4} + b_2 T_{m+3} + b_3 T_{m+2} + b_4 T_{m+1} + b_5 T_{m}=0
\ . 
\label{recursion}
\eeq
We also point out a related set of identities expressing the $a_j$ in
${\cal N}$ in terms of the $b_k$ in ${\cal D}$ and the Tutte
polynomials:
\beq
a_0 = T_1
\label{a0rel}
\eeq
\beq
a_1 = T_2 + b_1 T_1
\label{a1rel}
\eeq
\beq
a_2 = T_3 + b_1 T_2 + b_2 T_1
\label{a2rel}
\eeq
\beq
a_3 = T_4 + b_1 T_3 + b_2 T_2 + b_3 T_1
\label{a3rel}
\eeq
\beq
a_4 = T_5 + b_1 T_4 + b_2 T_3 + b_3 T_2 + b_4 T_1 \ .
\label{a4rel}
\eeq

As is evident from Eqs. (\ref{b1})-(\ref{b5}), 
\beq
b_j \ {\rm is \ invariant \ under} \ x \leftrightarrow y \ {\rm for \ each}
\ j = 1,...,5 \ . 
\label{bjxysym}
\eeq
Since this symmetry property obtains for each coefficient $b_j$, $j=1,...,5$,
it is also true for the full denominator of the generating function:
\beq
{\cal D}(S,x,y,z) \ {\rm is \ invariant \ under} \ x \leftrightarrow y \ .
\label{dxysym}
\eeq
However, the numerator polynomial ${\cal N}(S.x,y,z)$ does not have
this invariance (indeed, none of the $a_j$ is invariant under $x
\leftrightarrow y$), and hence it is also the case that the actual
Tutte polynomials $T(S_m,x,y)$ are not invariant under $x
\leftrightarrow y$.

Interestingly, although the graph $S_m$ is different from all of the
graphs considered in \cite{sdg}, the denominator polynomial ${\cal
  D}(S,x,y,z)$ is the same as the denominator polynomial for a strip
of the square lattice with width $L_y=2$ and self-dual boundary
conditions denoted as DBC1 calculated in \cite{sdg}.  
Thus, the $b_j$ with $1 \le j \le 5$ are
equal, respectively, to the $b_{Sj}$ in Eqs. (A21)-(A25) in
\cite{sdg}. However, the analysis carried out in Ref. \cite{sdg} 
was for the Tutte polynomial (equivalently, Potts partition
function) and chromatic polynomial, but not for the Jones polynomial,
so no analysis was performed in \cite{sdg} of the properties of
this denominator polynomial in the case $x=-t$ and $y=-1/t$ relevant
for Jones polynomials. 


\section{Analysis of ${\cal D}$ and $\lambda_j$ Terms} 
\label{lambda_section}

Since each coefficient function $b_j$ in ${\cal D}(S,x,y,z)$ is invariant
under the interchange $x \leftrightarrow y$, it follows that when one
carries out the substitutions $x=-t$ and $y=-1/t$ for the calculation of
the Jones polynomial, as specified in Eq. (\ref{vtrel}), each such $b_j$ has
the symmetry property
\beq
b_j \ {\rm is \ invariant \ under} \ t \leftrightarrow t^{-1} \
{\rm for \ each} \ j = 1,...,5 \ .
\label{bjtsym}
\eeq
Hence, the same symmetry holds for the full denominator function.
Defining
\beq
{\cal D}(S,z,t) \equiv {\cal D}(S,x,y,z)
\quad {\rm with} \quad x = -t, \quad y = -t^{-1} \ , 
\label{dentz}
\eeq
we have
\beq
{\cal D}(S,z,t) = {\cal D}(S,z,t^{-1}) \ . 
\label{densym}
\eeq
Performing the substitutions $x=-t$ and $y=-t^{-1}$ in the coefficients
$b_j(S,x,y)$, and denoting
\beq
b_j(S,x=-t,y=-t^{-1}) \equiv b_{jt} \ ,
\label{bjt}
\eeq
we obtain
\beqs
b_{1t} = -b_{4t} &=& -\frac{(t-1)^2(t^2-t+1)}{t^2} \cr\cr
                 &=& -4 + 3(t + t^{-1}) - (t^2 + t^{-2})
\label{b1t}
\eeqs
\beqs
b_{2t} = -b_{3t} &=& -\frac{(t-1)^2(t^2-t+1)^2}{t^3} \cr\cr
                 &=& 10 -8(t+t^{-1}) + 4(t^2+t^{-2}) -(t^3 +t^{-3}) 
\label{b2t}
\eeqs
\beq
b_{5t} = -1 \ .
\label{b5t}
\eeq
In Eq. (\ref{b1t}), the expression in the first line exhibits the
factorization properties of $b_{1t}$ and $b_{4t}$, while the expression
in the second line shows manifestly the invariance under the interchange
$t \leftrightarrow t^{-1}$, and similarly with Eqs. (\ref{b2t}) for
$b_{2t}$ and $b_{3t}$.

It is also useful to write the generating function in terms of the
inverse variable $\xi \equiv 1/z$. After the substitutions $x=-t$ and
$y=-t^{-1}$ above, the denominator of the generating function takes the
form
\beqs
    {\cal D}(S,\xi,t) &=& \xi^{-5}(\xi-1)
    \bigg [ 1+(1+b_{1t})\xi(1+\xi^2) + (1+b_{1t}+b_{2t})\xi^2 + \xi^4
    \bigg ] \cr\cr
    &=& \xi^{-5}(\xi-1)Q_s(\xi,t) \ , 
\label{dentxi}
\eeqs
where
\beqs
Q_s(\xi,t) &=& 1+(1+b_{1t})\xi(1+\xi^2) + (1+b_{1t}+b_{2t})\xi^2 + \xi^4
\cr\cr
&=& 1 + q_1 \xi(1+\xi^2) + q_2 \xi^2 + \xi^4
\label{qspol}
\eeqs    
with
\beq
q_1 = 1+b_{1t} = -3 + 3(t + t^{-1}) - (t^2 + t^{-2})
\label{q1}
\eeq
and
\beq
q_2 = 1+b_{1t}+b_{2t} = 7 -5(t+t^{-1})+3(t^2+t^{-2}) -(t^3+t^{-3}) \ .
\label{q2}
\eeq
The $t$-inversion symmetry 
\beq
Q_s(\xi,t) = Q_s(\xi,t^{-1})
\label{qs_sym}
\eeq
is manifest in Eqs. (\ref{q1})-(\ref{q2}), and we have indicated this
with the subscript $s$ for symmetry. This is accord with the 
$t$-inversion symmetry property for the full denominator function
given in Eq. (\ref{densym}).

Let us denote the roots of ${\cal D}(\xi,x,y)$ as a polynomial in
$\xi$ as $\lambda_i$ with $1 \le i \le 5$.  As discussed in earlier
works, e.g., \cite{bkw,w,strip,a}, the continuous accumulation locus
${\cal B}$ in the complex $t$ plane is determined as the solution to
the equality in magnitude of the dominant roots, i.e.,
\beq
{\cal B}: \quad |\lambda_i| = |\lambda_j| \ ,
\label{lamijdom}
\eeq
where these are larger in magnitude than other roots. ( Note that while
at generic points on ${\cal B}$, this equality only involves two
roots, at special points on ${\cal B}$ more than two dominant roots may
be degenerate.) Combining the inversion symmetry (\ref{densym}) and Eq.
(\ref{lamijdom}), we deduce that
\beq
    {\cal B} \ {\rm is \ invariant \ under \ the \ inversion} \ t \to
    \frac{1}{t} \ .
\label{B_inversion_sym}
\eeq

One of the roots of ${\cal D}(\xi,t)$ is the root of the
linear factor, $(\xi-1)$, which we denote as $\lambda_1$: 
\beq
\lambda_1 = 1 \ .
\label{lam1}
\eeq
The other four roots of ${\cal D}(\xi,t)$ depend on $t$.
Multiplying $Q_s(\xi,t)$ by $t^3$, we thus study the roots of the
quartic polynomial in $\xi$
\beqs
Q(\xi,t) &\equiv& t^3 Q_s(\xi,t) = t^3(\xi^4 + 1) +
t(-t^4+3t^3-3t^2+3t-1)(\xi^3 + \xi) \cr\cr
     &-& (t^3-t^2+2t-1)(t^3-2t^2+t-1)\xi^2 \ . 
\label{qpol}
\eeqs

Before giving the general solution to this quartic, we first discuss two
special cases, namely $t=1$ and $t=\pm i$.  At $t=1$ we have
\beq
Q(\xi,1) = 1+\xi+\xi^2+\xi^3+\xi^4 = \frac{\xi^5-1}{\xi-1} \ .
\label{qpol_t1}
\eeq
The roots of $Q(\xi,1)$ are thus the fifth roots of unity with the
exception of $\xi=1$. Together with the root $\xi=1$ from the linear
factor in Eq. (\ref{dentxi}), these thus comprise the five roots of
${\cal D}(\xi,t)$ in $\xi$ for $t=1$. It follows that at $t=1$, all
five roots of ${\cal D}(\xi,t)$ have the same magnitude, namely
unity. This shows that this point, $t=1$, is on the locus ${\cal B}$
and is a multiple point, in the technical terminology of algebraic
geometry \cite{hartshorne}, i.e., a
point at which different branches on ${\cal B}$ cross.  As will be
shown below, the branches that intersect at $t=1$ are comprised of a
(self-conjugate) arc of the unit circle and a finite
inversion-symmetric line segment along the (positive) real axis in the
$t$ plane. With $\xi=\lambda_1=1$ as given in Eq. (\ref{lam1}), we
denote the four roots of $Q(\xi,t)$ as $\lambda_j$, with $2 \le j \le 5$.
At $t=1$ we have
\beqs
t=1 \ \Rightarrow \ \lambda_{j=2,3} &=& e^{\pm 2\pi i/5} =
\frac{1}{4}\bigg [-1+\sqrt{5} \pm i\sqrt{2(5+\sqrt{5})} \
  \bigg ] \cr\cr
               &=& 0.309017 \pm 0.9510565i 
\label{lam_t1_23}
\eeqs
and
\beqs
t=1 \ \Rightarrow \ \lambda_{j=4,5} &=& e^{\pm 4\pi i/5} =
\frac{1}{4}\bigg [-1-\sqrt{5} \pm i\sqrt{2(5-\sqrt{5})} \
  \bigg ] \cr\cr
            &=& -0.809017 \pm 0.587785i \ .
\label{lam_t1_45}
\eeqs
At $t=i$, we have 
\beqs
Q(\xi,i) &=& -i(1-\xi+\xi^2-\xi^3+\xi^4) =
-i\Bigg (\frac{\xi^5+1}{\xi+1}\Bigg ) \cr\cr
&=& -\frac{i}{4} \Big [ 2\xi^2 + (1-\sqrt{5})\xi + 2 \ \Big ]
                \Big [2\xi^2  - (1+\sqrt{5})\xi + 2 \ \Big ] \ .
\label{qpol_ti}
\eeqs
The roots of $Q(\xi,i)$ are thus the fifth roots of $-1$ with the
exception of $\xi=-1$:
\beqs
t=i \ \Rightarrow \ \lambda_{4,5} &=& e^{\pm \pi i/5} =
  \frac{1}{4}\bigg [1+\sqrt{5} \pm i\sqrt{2(5-\sqrt{5})} \
  \bigg ] \cr\cr
&=& 0.809017 \pm 0.587785i
\label{lam_ti_45}
\eeqs
and
\beqs
t=i \ \Rightarrow \ \lambda_{2,3} &=& e^{\pm 3\pi i/5} =
  \frac{1}{4}\bigg [1-\sqrt{5} \pm i\sqrt{2(5+\sqrt{5})} \
  \bigg ] \cr\cr
&=& -0.309017 \pm 0.9510565i \ .
\label{lam_ti_23}
\eeqs
Together with the root of ${\cal D}$ at $\xi=1$,
this again means that at $t=\pm i$ all five roots of ${\cal D}$ have
equal magnitude, equal to unity. In turn, this implies that the points
$t=\pm i$ are multiple points on ${\cal B}$. 

To express the general solutions for the roots of the quartic polynomial
$Q(\xi,t)$ in $\xi$, we first define some auxiliary polynomials.  Let
\beq
P_a = t^4 -3t^3 +3t^2-3t+1
\label{pola}
\eeq
\beq
P_b = t^8-4t^7+9t^6-14t^5+11t^4-14t^3+9t^2-4t+1
\label{polb}
\eeq
\beq
R_1 = t^8-2t^7+3t^6-4t^5+9t^4-4t^3+3t^2-2t+1 \ .
\label{R1}
\eeq
Note the symmetry relations 
\beq
P_a = t^4 P_a(t \to t^{-1})
\label{pasym}
\eeq
\beq
P_b = t^8 P_b(t \to t^{-1})
\label{pbsym}
\eeq
\beq
R_1 = t^8 R_1(t \to t^{-1})  \ .
\label{r1sym}
\eeq
The solutions for $\xi$ to the quartic polynomial equation
$Q(\xi,t)=0$ are then as follows, where we use a self-evident
labelling with subscripts $p$ and $m$ standing for the plus or minus
signs in front of the three square roots:
\beqs
\lambda_{ppp} &=& \frac{1}{4t^2}\bigg [ P_a + \sqrt{R_1} +
  \sqrt{2(P_b + P_a\sqrt{R_1})} \ \bigg ]   
\label{lam_ppp}
\eeqs
\beqs
\lambda_{pmp} &=& \frac{1}{4t^2}\bigg [ P_a + \sqrt{R_1} -
  \sqrt{2(P_b + P_a\sqrt{R_1})} \ \bigg ]   
\label{lam_pmp}
\eeqs
\beqs
\lambda_{mpm} &=& \frac{1}{4t^2}\bigg [ P_a - \sqrt{R_1} +
  \sqrt{ 2(P_b - P_a\sqrt{R_1}) } \  \bigg ]   
\label{lam_mpp}
\eeqs
\beqs
\lambda_{mmm} &=& \frac{1}{4t^2}\bigg [ P_a - \sqrt{R_1} -
  \sqrt{ 2(P_b - P_a\sqrt{R_1}) } \ \bigg ]  \ . 
\label{lam_mmm}
\eeqs
We remark that a generic solution for the roots of a quartic
polynomial involves cube roots as well as square roots.  The reason
that the solution for $\xi \equiv \lambda$ to the equation
$Q(\xi,t)=0$ only involves square roots and nested square roots, but
no cube roots, can be traced to the special properties of
Eq. (\ref{qpol}), including the fact that in $Q(\xi,t)$ the
coefficients of the constant and $\xi^4$ terms are the same and,
separately, the coefficients of the $\xi^3$ and $\xi$ terms in
$Q(\xi,t)$ are the same.


\section{The Locus ${\cal B}$ }
\label{locus_section}

In this section we study the continuous accumulation set ${\cal B}$ of
the zeros of $V(H_r,t)$ in the limit $r \to \infty$, which can thus be
written as ${\cal B}(H_\infty)$.  Since the set of zeros of a Jones
polynomial is invariant under complex conjugation, it follows that
this is also true of the locus ${\cal B}$, i.e., ${\cal B} = {\cal
  B}^*$.  An important property of ${\cal B}$ for $H_\infty$ is that
it is symmetric under the inversion $t \to 1/t$, as derived in
Eq. (\ref{B_inversion_sym}). For finite $m$, many zeros of $V(H_r,t)$
lie close to the asymptotic locus ${\cal B}$.  In
Fig. \ref{vhzeros_fig} we show an illustrative set of zeros of
$V(H_r,t)$ with $r=402$ in the complex $t$ plane.

\begin{figure}[htbp]
  \begin{center}
  \includegraphics[height=18cm,width=10cm]{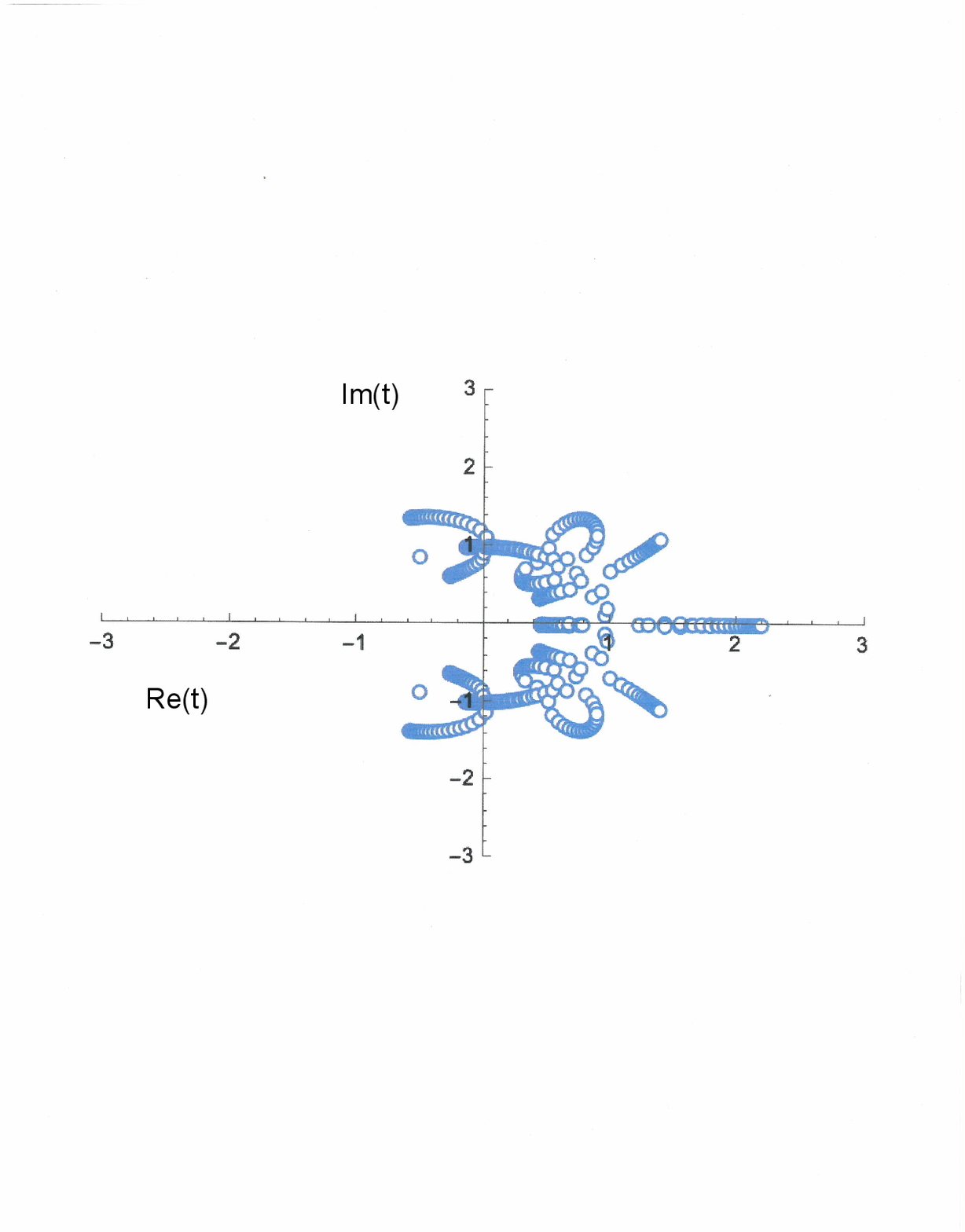}
  \end{center}
  \caption{Illustrative plot of zeros of $V(H_r,t)$ with $r=402$ in the
    complex $t$ plane.}
\label{vhzeros_fig}
\end{figure}

We find that this locus ${\cal B}$ for $H_\infty$ is the union of
several parts, including

\begin{enumerate}

\item

  A finite line segment ${\cal L}$ on the positive real axis.
  This line segment is inversion-symmetric and thus necessarily passes through
  the point $t=1$. 

\item
  An arc ${\cal C}$ of the unit circle passing through $t=1$ and
  $t=\pm i$ and extending slightly into the second and third quadrants
  of the $t$ plane.

\item
  A complex-conjugate pair of leftwardly concave horseshoe-shaped arcs
  ${\cal H}$ and ${\cal H}^*$ that intersect the above-mentioned arc
  of the unit circle at $\pm i$ and extend into the second and
  third quadrants of the $t$ plane.

\item
  A complex-conjugate pair of arcs ${\cal A}$ and ${\cal A}^*$ in the
  first and fourth quadrants that cross the arc ${\cal C}$ of the unit circle.

\item

  An additional complex-conjugate pair of loop-like curves straddling the arc
  ${\cal C}$ of the unit circle in the first and fourth quadrant.

\end{enumerate}
The zeros of $V(H_r,t)$ for finite $r$ and hence $m=m(r)$ in
Fig. \ref{vhzeros_fig} can be seen to approximate these parts of the
asymptotic locus ${\cal B}$.  These zeros also include a complex-conjugate
pair of isolated zeros of unit magnitude very near to
\beq
\{ t_{isol.}, \ t_{isol.}^* \ \}
= e^{\pm 2 i \pi/3} = \frac{1}{2}(-1 \pm i \sqrt{3} \ ) \ .
\label{tisol}
\eeq
These zeros do not, in general, occur exactly at $e^{\pm 2 i \pi/3}$
since if they did, then the polynomial
$(t-e^{2i\pi/3})(t-e^{-2i\pi/3}) = t^2 + t + 1$ would have be a factor
of $V(H_r,t)$ and this is not the case, as can be seen in the explicit
results in Eqs. (\ref{vm1})-(\ref{vm3}). This occurrence of a complex-conjugate pair of zeros very
near to $t = e^{\pm 2 i \pi/3}$ was also observed for a different
family of repeating chain knots in (section 5 of) \cite{ww}. The 
finite-$r$ calculations carried out in \cite{ww} were consistent with
the conclusion that in the limit of infinitely many crossings, this pair
would coincide with $t=e^{\pm 2i\pi/3}$, and this is
also true of our calculation of zeros of $V(H_r,t)$.  Interestingly,
the points $t=e^{\pm 2i\pi/3}$ are among the exceptions to the
$\sharp{\rm P}$ complexity in the calculation of the Jones polynomial of
a knot or link \cite{jvw}.

Concerning the density of zeros for finite $r$, the decrease in this
density in the vicinity of the multiple point at $t=1$ in
Fig. \ref{vhzeros_fig} is similar to what was found in \cite{jz} for
the zeros of $V(B_m,t)$ in the vicinity of $t=1$ (see Fig. 4
of \cite{jz}). Indeed, similar decreases in the densities of zeros of
statistical mechanical partition functions in temperature-dependent
Boltzmann variables (Fisher zeros) and zeros of chromatic polynomials
(chromatic zeros) in the vicinity of multiple points of algebraic
curves forming respective accumulation loci ${\cal B}$ were found in a
number of studies; see, e.g., Fig. 1 in \cite{only}, Figs. 2(a) and 6(a)
in \cite{nec}, Figs. 4 and 7 in \cite{a}, and Fig. 7 in \cite{s3a},
for example.

In studies of continuous accumulation loci of chromatic zeros for
families of graphs in the limit of infinitely many vertices, it was
shown that endpoints of arcs in the complex plane generically arise as
zeros in various roots, e.g., square roots in \cite{strip,a}.  The
situation is similar here.  The horseshoe-shaped arcs ${\cal H}$ and
${\cal H}^*$, and the arcs ${\cal A}$ and ${\cal A}^*$ together have
eight endpoints, and these are given as the eight zeros of the
degree-8 polynomial $R_1$ in Eq. (\ref{R1}) that occurs in the quartic
roots (\ref{lam_ppp})-(\ref{lam_mmm}). We denote the endpoint of the
arc ${\cal A}$ that is outside of the unit circle in the $t$ plane as
$t_{{\cal A},o}$ and the endpoint that is inside of this unit circle
as $t_{{\cal A},i}$, where the subscripts $o$ and $i$ stand for
``outside'' and ``inside''. Similarly, we label the endpoint of the
horseshoe-shaped arc in the second quadrant that has magnitude greater
than (resp. less than) $|t|=1$ as $t_{{\cal H},o}$ (resp.  $t_{{\cal
    H},i}$). The $t \to 1/t$ inversion symmetry of ${\cal B}$ implies
\beq
         |t_{{\cal A},i}| =
\frac{1}{|t_{{\cal A},o}|}
\label{arc1_endpoint_sym}
\eeq
and
\beq
         |t_{{\cal H},i}| =
\frac{1}{|t_{{\cal H}_o}|} \ . 
\label{arc2_endpoint_sym}
\eeq
Calculating the eight zeros of $R_1$ we find
\beq
\{ t_{{\cal A}_o}, \
   t_{{\cal A}_o}^* \ \} = 1.398781 \pm 1.091186i
\label{arc1_outer_endpoints}
\eeq
\beq
\{ t_{{\cal A}_i}, \
   t_{{\cal A}_i}^* \ \} =  0.444442 \pm 0.346708i
\label{arc1_inner_endpoints}
\eeq
\beq
\{ t_{{\cal H}_o}, \
   t_{{\cal H}_o}^* \ \} = -0.579679 \pm 1.365109i
\label{arc2_outer_endpoints}
\eeq
\beq
\{ t_{{\cal H}_i}, \
   t_{{\cal H}_i}^* \ \}  = -0.263544 \pm 0.620631i
 \ .
\label{arc2_inner_endpoints}
\eeq

Further insight into the roots of $Q(\xi,t)$ and hence into the
properties of the locus ${\cal B}$ can be gained by an analysis of the
discriminant of the equation $Q(\xi,t)=0$, which we denote as ${\rm
  Disc}(Q)$. Recall that this discriminant vanishes if two of the
roots of $Q(\xi,t)$ are equal \cite{disc}.  Hence, one may solve for
values of $t$ where ${\rm Disc}(Q)=0$ and then determine the subset of
these values such that the corresponding values of $\lambda$ are
maximal.  These are then on ${\cal B}$.  We calculate the discriminant
to be, in factorized form,
\beq
    {\rm Disc}(Q) = t^4 D_1 D_2 D_3 D_4^2 \ , 
\label{discriminant}
\eeq    
where
\beq
D_1 = t^3-3t^2+2t-1
\label{disc1}
\eeq
\beq
D_2 = t^3-2t^2+3t-1
\label{disc2}
\eeq
\beq
D_{12} \equiv D_1D_2 = t^6-5t^5+11t^4-15t^3+11t^2-5t+1
\label{disc12}
\eeq
\beq
D_3 = t^6-t^5-t^4-3t^3-t^2-t+1
\label{disc3}
\eeq
\beq
D_4 = t^8 -2t^7+3t^6-4t^5+9t^4-4t^3+3t^2-2t+1 \ .
\label{disc4}
\eeq
As with other relevant polynomials, we note the symmetry relations under
the inversion $t \to t^{-1}$:
\beq
D_1 = -t^{-3}D_2(t \to t^{-1}), \quad\quad  D_2 = -t^{-3}D_1(t \to t^{-1})
\label{disc1_disc2_sym}
\eeq
\beq
D_{12} = t^6 D_{12}(t \to t^{-1})
\label{ddisc12_sym}
\eeq
\beq
D_3 = t^6 D_3(t \to t^{-1})
\label{disc3_sym}
\eeq
\beq
D_4 = t^8 D_4(t \to t^{-1}) \ . 
\label{disc4_sym}
\eeq
These symmetry relations imply that the set of zeros of the discriminant
${\rm Disc}(Q)$ is invariant under the inverse $t \to t^{-1}$. 

Two of the zeros of ${\rm Disc}(Q)$ determine the
endpoints of the line segment on ${\cal B}$, which are larger than $t=1$
and smaller than $t=1$ on the (positive) real axis. For uniformity with our
previous labelling, we denote these as outer and inner endpoints,
$t_{{\cal L},o}$ and $t_{{\cal L},i}$. 
These are given by the two real zeros of the 
degree-6 polynomial factor $D_3$, namely 
\beq
t_{{\cal L},o} = 2.1956467
\label{lse_outer}
\eeq
and
\beq
         t_{{\cal L}_i} =
\frac{1}{t_{{\cal L},o}} = 0.45544667 \ .
\label{lse_inner}
\eeq
(The other four zeros of $D_3$ are at complex values of $t$.)  To show
that these points (\ref{lse_outer}) and (\ref{lse_inner}) are, in
fact, on ${\cal B}$, we observe that at each of these values of $t$,
the roots $\xi=\lambda$ of $Q(\xi,t)$ are degenerate in magnitude and
are dominant. We find that at each of these endpoints,
the dominant roots of $Q(\xi,t)$ consist of $\lambda=1$ with multiplicity 2 and
a complex-conjugate pair of complex $\lambda$'s of magnitude 1, namely
$\lambda = -0.962492 \pm 0.271310i$.

We denote the endpoints of the arc of the unit circle as
$\{t_{{\cal C}_e},t_{{\cal C}_e}^* \} = e^{\pm i \theta_{{\cal C}_e}}$, 
where the subscript $e$ stands for
``endpoint''. These values are determined by the condition that the
argument of the square root in $\lambda_{ppm}$ and $\lambda_{pmm}$
vanishes, i.e., such that $P_b-P_a \sqrt{R_1}=0$. To determine this,
we consider solutions to the equation $P_b^2-P_a^2 R_1 = 0$.  The
polynomial $P_b^2-P_a^2 R_1$ factorizes according to
\beq
P_b^2-P_a^2 R_1 = 4t^2 D_1 D_2 D_3 \ , 
\label{roottpol}
\eeq
where $D_j$, $j=1,2,3$, are factors in the discriminant ${\rm Disc}(Q)$
and were given in Eqs. (\ref{disc1})-(\ref{disc3}). 
The complex-conjugate pair of endpoints of the circular arc on ${\cal B}$ in,
respectively, the second and third quadrants occur at a pair of zeros of 
$D_3$, namely
\beq
\{ t_{{\cal C}_e}, \ t_{{\cal C}_e}^* \} = -0.136945 \pm 0.990579i =
        e^{\pm i\theta_{{\cal C}_e}} \ ,
\label{tcae}
\eeq
with
\beq
\theta_{{\cal C}_e} = 97.8711^\circ \ .
\label{theta_cae}
\eeq

We remark that the locus ${\cal B}$ for $H_\infty$ has two features in
common with the locus ${\cal B}$ calculated in Ref. \cite{jz} for the
$m \to \infty$ limit of the family of associated graphs denoted $B_m$,
namely the presence of a real line segment and an arc of the unit circle,
although (i) the endpoints of this line segment and arc are different for
$H_\infty$ and $B_\infty$; and (ii) for $B_\infty$ they comprise the totality
of the locus ${\cal B}$, whereas for $H_\infty$ they comprise only part
of the locus.  Explicitly, the locus ${\cal B}$ for $B_\infty$ was given 
in Eq. (5.13) of \cite{jz} as 
\beq
    {\cal B}(B_\infty): \quad
\bigg \{ \ t=e^{i\theta} \ , \quad -\frac{2\pi}{3} \le \theta \le
      \frac{2\pi}{3} \bigg \} \quad \cup \quad
 \bigg \{ \ \frac{3-\sqrt{5}}{2} \le t \le \frac{3+\sqrt{5}}{2} \ \bigg \} \ .
\label{bvb}
\eeq
This locus ${\cal B}(B_\infty)$ is also invariant under the
inversion map $t \to t^{-1}$, just as ${\cal B}(H_\infty)$ is.
Numerically, $(1/2)(3-\sqrt{5} \ ) = 0.381966$ and $(1/2)(3+\sqrt{5}
\ ) = 2.618034$, so the line segment on ${\cal B}(B_\infty)$ is
slightly longer than the one here, with endpoints given in Eqs.
(\ref{lse_inner}) and (\ref{lse_outer}). Similarly, the endpoints of the
circular arc on ${\cal B}$ for $B_\infty$ occur at $\theta= \pm 2\pi/3
= \pm 120^\circ$, which extend slightly farther into the second and
third quadrant than the endpoints of the circular arc in the locus
${\cal B}$ for $H_\infty$ analyzed here, as given in Eqs. (\ref{tcae}) and
(\ref{theta_cae}). Note that the points $t=e^{\pm 2i \pi/3}$
play a special role for ${\cal B}(B_\infty)$, but not as limits of
isolated zeros as $r \to \infty$; instead, they are endpoints of the
circular arc on ${\cal B}(B_\infty)$. In general, the locus ${\cal B}$
for $H_\infty$ is substantially more complicated than the locus ${\cal
  B}$ for $B_\infty$ in \cite{jz}. This is understandable, since
$V(B_r,t)$ involves a smaller number of terms in Eq. (\ref{vlsum}),
namely $N_{B_r,\lambda}=3$, as compared with $N_{H_r,\lambda}=5$ here.


\section{Conclusions }
\label{conclusion_section}

In this paper we have calculated Jones polynomials $V(H_{r(m)},t)$ for
a family of alternating knots and links with arbitrarily many
crossings $r(m)=4m+2$, $m \in {\mathbb Z}_+$, by computing the Tutte
polynomials $T(G_+(H_{r(m)}),x,y)$ for the associated graphs
$G_+(H_{r(m)})=S_m$.  Our results further elucidate how, with this
method, one can calculate Jones polynomials for an infinite family of
this type in a systematic manner, circumventing the generic
property that the calculation of the Jones polynomial of knots and
links becomes intractably difficult for sufficiently many
crossings. We have studied the zeros of $V(H_{r(m)},t)$ in the complex
$t$ plane and the accumulation set ${\cal B}(H_\infty)$. These results
involve a fruitful confluence of knot theory, graph theory, and
complex analysis. 

\bigskip
\bigskip


{\bf Acknowledgments}

This research was supported in part by the NSF grant PHY-22-10533.  RS
is grateful to S.-C. Chang for the earlier collaboration on Ref. \cite{jz}:


\bigskip
\bigskip

{\bf Declarations}: (1) The authors have no financial or
non-financial conflicts of interest relevant to this article; (2)
concerning data accessibility, this article is theoretical, and
relevant data are included herein.


\bigskip
\bigskip

\begin{thebibliography}{99}
  
\bibitem{rolfsen}
Rolfsen, D.: Knots and Links. Publish or Perish, Berkeley (1976) 

\bibitem{bz}
Burde, G. and Zieschang, H.: Knots. de Gruyter, New York (1985)

\bibitem{kauffman}
Kauffman, L.: Knots and Physics. World Scientific, Singapore (1991)

\bibitem{wu_knots}
Wu, F. Y.: Knot theory and statistical mechanics. Rev. Mod. Phys. {\bf 64}, 
1099-1131 (1992)

\bibitem{welsh}
Welsh, D. J. A.: Complexity: Knots, Colourings, and Counting. Cambridge
University Press, Cambridge (1993)

\bibitem{lickorish}
Lickorish, W. B. R.: An Introduction to Knot Theory. Springer, Berlin (1997)

\bibitem{bollobas}
Bollob\'as, G.: Modern Graph Theory. Springer, New York (1998)
  
\bibitem{jones85}
Jones, V. F. R.: A polynomial invariant for links via von
Neumann algebras. Bull. Am. Math. Soc. {\bf 12}, 103-112 (1985)

\bibitem{jones87}
Jones, V. F. R.: Hecke algebra representations of braid
groups and link polynomials. Ann. Math. {\bf 126}, 335-388 (1987)

\bibitem{jones89}
Jones, V. F. R.: On knot invariants related to some statistical
mechanical models, Pacific J. Math. {\bf 137}, 311-334 (1989)

\bibitem{thistle87}
Thistlethwaite, M. B.: A spanning tree expansion of the Jones polynomial.
Topology {\bf 26}, 297-309 (1987)

\bibitem{jvw}
Jaeger, F., Vertigan, D. L., and Welsh, D. J. A.: On the
  computational complexity of the Jones and Tutte polynomials.
  Math. Proc. Camb. Phil. Soc. {\bf 108}, 35-53 (1990)

\bibitem{birman1993}
Birman, J.: New points of view in knot theory. Bull. Am. Math. Soc.
  {\bf 28}, 253-287 (1993) 

\bibitem{ww}
  Wu, F. Y. and Wang, J.: Zeroes of the Jones polynomial.
  Physica A {\bf 296}, 483-494 (2001)

  
\bibitem{jz}
Chang, S.-C. and Shrock, R.:
Zeros of Jones polynomials for families of knots and links. 
Physica A {\bf 301}, 196-218 (2001)


\bibitem{jin_zhang2003}
Jin, X. and F. Zhang, F.: Zeros of the Jones polynomials for families of
  pretzel links. Physica A {\bf 328}, 391-408 (2003) 

\bibitem{jin_zhang2010}
Jin, X. and Zhang, F.: Zeros of the Jones polynomial for multiple
  crossing-twisted links. J. Stat. Phys. {\bf 140}, 1054-1064 (2010) 

\bibitem{dong_dense}
Jin, X., Zhang, F., Dong, F., and Tay, E. G.: 
  Zeros of the Jones polynomial are dense in the complex plane.
  Electronic J. Combin. {\bf 17}, R94 (2010)

\bibitem{dong_jin2015}
  Dong, F. and Jin, X.: Zeros of Jones polynomials of graphs.
  Electronic J. Combin. {\bf 22}, P3.23 (2015) 
  
\bibitem{tutte1}
Tutte, W. T.: A Contribution to the theory of chromatic polynomials.
  Canadian. J. Math. {\bf 6}, 80-81 (1954)
  
\bibitem{tutte2}
Tutte, W. T.: On dichromatic polynomials. J. Combin. Theory {\bf 2},
  301-320 (1967)

\bibitem{a}
Shrock, R.: Exact Potts model partition functions for ladder graphs.
Physica A {\bf 283}, 388-446 (2000)

\bibitem{bkw}
  Beraha, S., Kahane, J., and Weiss, H.: Limits of zeros
  of recursively defined families of polynomials. In: Rota,
  G. C. (ed.), Studies in Foundations and Combinatorics, Advances in
  Mathematics and Supplementary Studies, vol. 1, pp. 213-232. Academic
  Press, New York (1978)

\bibitem{w}
Shrock, R. and S.-H. Tsai, S.-H.:
Asymptotic limits and zeros of chromatic polynomials and ground state
entropy of Potts antiferromagnets, Phys. Rev. E {\bf 55}, 5165-5179 (1997)

  
\bibitem{strip}
Ro\v{c}ek, M., Shrock, R., and Tsai, S.-H.:
Chromatic polynomials for families of strip graphs and their asymptotic      
limits. Physica A {\bf 252}, 505-546 (1998)
  
\bibitem{bds}
Biggs, N. L., Damerill, R. M., and Sands, D. A.:
  Recursive families of graphs. J. Combin. Theory {\bf 12}, 123-131 (1972)


\bibitem{sdg}
Chang, S.-C. and Shrock, R.:
Complex-Temperature Phase Diagrams for the $q$-State Potts Model on
Self-Dual Families of Graphs and the Nature of the $q \to \infty$ Limit.
Phys. Rev. E {\bf 64}, 066116 (2001).
  
\bibitem{hartshorne}
Hartshorne, R.: Algebraic Geometry, Springer, New York ( 1977)

\bibitem{only}
Matveev, V. and Shrock, R.:
Complex-temperature properties of the 2D Ising model for nonzero magnetic    
field, Phys. Rev. E {\bf 53}, 254-267 (1994)
  
\bibitem{nec}
Shrock, R. and Tsai, S.-H.:
Ground state entropy of Potts antiferromagnets on cyclic                     
polygon chain graphs, J. Phys. A {\bf 32}, 5053-5070 (1999)

\bibitem{s3a}
Chang, S.-C. and Shrock, R.:  
Exact Potts model partition functions on wider arbitrary-length              
strips of the square lattice. Physica A {\bf 296}, 234-288 (2001)


\bibitem{disc}
Gelfand, I. M., Kapranov, M. M., and Zelevinsky, A. V.: Discriminants,
    Resultants, and Multidimensional Determinants, Birkhauser, Boston (1994)
  
\end{thebibliography}
\end{document}